# Comprehensive and User-Analytics-Friendly Cancer Patient Database for Physicians and Researchers

Ali Firooz
*College of Engineering and Computing*
*University of South Carolina*
Columbia, SC, USA
ali.firooz@sc.edu

Avery T. Funkhouser
*School of Medicine Greenville*
*University of South Carolina*
Greenville, SC, USA
averytf@email.sc.edu

Julie C. Martin
*Prisma Health Cancer Institute*
Greenville, SC, USA
julie.martin@prismahealth.org

W. Jeffery Edenfield
*Prisma Health Cancer Institute*
Greenville, SC, USA
jeffery.edenfield@prismahealth.org

Homayoun Valafar
*College of Engineering and Computing*
*University of South Carolina*
Columbia, SC, USA
homayoun@cse.sc.edu

Anna V. Blenda
*School of Medicine Greenville*
*University of South Carolina*
*Prisma Health Cancer Institute*
Greenville, SC, USA
ablenda@greenvillemed.sc.edu

*Abstract*—Nuanced cancer patient care is needed, as the development and clinical course of cancer is multifactorial with influences from the general health status of the patient, germline and neoplastic mutations, co-morbidities, and environment. To effectively tailor an individualized treatment to each patient, such multifactorial data must be presented to providers in an easy-to-access and easy-to-analyze fashion. To address the need, a relational database has been developed integrating status of cancer-critical gene mutations, serum galectin profiles, serum and tumor glycomic profiles, with clinical, demographic, and lifestyle data points of individual cancer patients. The database, as a backend, provides physicians and researchers with a single, easily accessible repository of cancer profiling data to aid-in and enhance individualized treatment.

Our interactive database allows care providers to amalgamate cohorts from these groups to find correlations between different data types with the possibility of finding "molecular signatures" based upon a combination of genetic mutations, galectin serum levels, glycan compositions, and patient clinical data and lifestyle choices. Our project provides a framework for an integrated, interactive, and growing database to analyze molecular and clinical patterns across cancer stages and subtypes and provides opportunities for increased diagnostic and prognostic power.

## I. Introduction

### A. Cancer Stats and Why New Strategies are Needed

Cancer is a significant burden and is the second leading cause of death after heart disease in the United States. In 2022 alone, the number of new cancer cases is projected to be close to 2 million with 600,000 deaths [1]. It is predicted that by the year 2030, the national costs of cancer care will balloon to a quarter of a trillion dollars [2]. Strategies to enable enhanced secondary prevention of cancer – additional ways to measure and detect cancer at earlier stages – are profoundly needed, as cancer prognoses are worse the later the cancer is found [3].

### B. Different Approaches to Cancer Screening

Glycomics has recently shown distinct potential in the area of cancer detection. Glycomics is the systematic study of all glycan (i.e., sugar) structures of a certain cell type or tissue. It is a promising strategy aimed at differentiating and diagnosing cancer [4]. Additionally, alterations in glycosylation have been correlated with tumor initiation, progression, and metastasis [5]. The tumor-associated glycans or glycoproteins are secreted or membrane-shed in the blood and can ultimately be used as markers for tumor presence [6].

In addition to using glycans as markers of cancer, proteins that interact with glycan epitopes can also serve as possible cancer signals. Galectins are a subset of the lectin family of proteins and share a high affinity β-galactoside binding domain which allows them to interact with other proteins by binding to glycosylation sites [7]. Expression of several galectins is known to be dysregulated in tumor cells and to have a variety of contributory and inhibitory roles on tumor cells and cancer environment [8–22]. As some of these galectins are significantly dysregulated in cancer, they have value as markers of cancer detection and progression [23].

Using galectin and glycomic profiling of cancer patients would be similar to other profiling methods already used in clinical practice. For instance, genomic and epigenomic profiling has shown that specific DNA methylation which have led to better clinical outcomes for glioblastoma multiforme [24]. Metabolomic profiling, based on tumor cells' complex metabolic requirements, has also been used as a hallmark for cancer, such as upregulation of glucose and glutamine metabolism and convergence of many metabolic pathways on glucose and glutamine to provide a more nutritious environment for tumor growth [25].

We predict galectin and glycomic profiles can be similarly used as other benchmarks for cancer progression, and potentially diagnosis and treatment. Furthermore, since genetic profile and a patient's individual lifestyle factors also play a role in oncogenesis, we predict a "stage signature" or "molecular subtype signature" can be generated based upon a combination of data such as glycan signatures, galectin serum levels, genetic mutations, clinical data, lifestyle choices, etc.

### C. Background Importance of Using Multifactorial Data

A model created by Zuo et. al. integrated gene expression, gene mutation, and chemical structure, an approach that showed to greatly enhance predictive performance of cancer drug responses [26]. Similarly, a model by Xie et al. demonstrated how multifactorial input data can be used to demonstrate the mechanisms of resistance to immunotherapy [27]. These studies demonstrate the utility of combining unique data types to discover correlations that can be implemented for care. Precision medicine has recently been at the forefront of new medical approaches, and it has been shown that attempting new discoveries as such must take advantage of new technologies and computational methodologies [28, 29]. Furthermore, it has been recommended that health professionals should lend a greater emphasis on combining biological, social, and environmental metrics to develop precision interventions to patients and the population [30].

### D. Why Combining Data in a Convenient Manner is Critical

"Big data" in healthcare refers to the exponentially increasing amount of health data that is collected and stored. The advent of this much information being at the access of providers' fingertips demonstrates a goldmine of potential that can be used for medical discovery [31]. However, one caveat to using large amounts of data and new technological approaches is that presentation in easy-to-use manner for physicians is of utmost importance, otherwise only marginal improvements have been demonstrated [32–35]. West et al. further suggest that there is significant potential for discovery using the advent of increasing healthcare data so long as the data are effectively managed, presented, and visualized [36].

### E. Aim

The goal of our project was the enhancement of the cancer patient clinical database with molecular data, for better diagnosis and clinical trials/treatment, and additional revenue. Our interactive database allows providers to amalgamate cohorts from their specific patient population to find correlations between different data types with the possibility of, for example, describing a "stage signature" based upon a combination of glycan signatures, galectin serum levels, genetic mutations, and lifestyle choices. Our project demonstrates, to the best of our knowledge, the first time such an interactive and patient-centered database has been made. We believe this utility, such as the ability to analyze patterns (signatures) across stages in our database, will provide an opportunity for increased diagnostic and prognostic power and perhaps change the outlook of what is possible with current patient data at point-of-care.

## II. MATERIALS & METHODS

### A. Data Categories

#### 1) Reported Data

*a) Patient Data* – Demographic, lifestyle, and disease-specific data were collected from the Prisma Health Cancer Institute Biorepository (PHCIB). The reported information included age, race, sex, smoking status, as well as tumor information such as TNM staging, grade, histology, and biopsy site.

*b) Samples* – The samples obtained from the PHCIB included blood serum from cancer patients and healthy control group, as well as malignant tissue and surrounding benign tissue from surgical resections performed on cancer patients.

#### 2) Produced Data

*a) Galectins* – The serum concentration of galectins -1, -3, -7, -8 and -9 were determined as previously described [23].

*b) Glycans* – Mass spectra were acquired by Bruker ultrafleXtreme MALDI TOF/TOF (Bruker, Billerica, MA). Data were collected from a mass range of 500 to 5,000 Daltons with positive mode. Following acquisition, masses were automatically identified with detection limits for peaks at S/N>3. This allowed the derivation of serum and tissue glycan composition levels of healthy controls and patients with cancer. Glycomic profiles were then created for each patient.

*c) Gene Mutations* – Genetic mutation data for 74 patients were provided by the PHCIB. The mutation status of 50 cancer-critical genes was derived by the Ion Ampliseq Cancer Hotspot Panel v2 as previously described [37]. The Ampliseq panel covers 50 oncogene and tumor suppressor genes, amplifying 207 amplicons covering approximately 2,800 COSMIC mutations. These 50 genes were then sorted into eleven groups of different cellular pathways according to the gene product role.

### B. Database

#### 1) Storage

The patient data were then incorporated into a Relational Database Management System (RDBMS) using Structured Query Language (MySQL) server version 8.0.31 on a Linux machine running Ubuntu 20.04.

The most widely used database system worldwide is RDBMS, as it provides a dependable method of storing and retrieving large amounts of data while offering a combination of system performance and ease of implementation. In addition, RDBMS bases the structure of its data on the Atomicity, Consistency, Isolation, and Durability (ACID) model to ensure and maintain the security, accuracy, integrity, and consistency of the data [38]. This implementation makes the database scalable for future deployment on a more resourceful server such as IFESTOS (ifestos.cse.sc.edu) or Amazon Web Services (AWS).

The primary database relation is the basic patient information identified by anonymous IDs implemented as the primary key. However, due to the complication of the data being stored in the database, several relations were in need for a more robust primary key to be uniquely identifiable. Therefore, a combination of attributes acting as the primary key were chosen. Furthermore, relations such as "CancerInfo", holding patient cancer information, are all connected to the primary relation "PersonInfo" by utilizing foreign keys that reference the primary key attribute. Additionally, several triggers were incorporated in the database to validate the data quality.

*2) Interface*

The initial interface to the database was facilitated through an intuitive web application that serves as a database administration tool, "phpMyAdmin", version 4.9.5 residing on top of "Nginx" server version 1.18.0.

To further ease the process of importing new data and minimize human mistakes, a Python script was developed. Even though MySQL offers an option to import Excel files directly into the database, doing so will pose several limitations on the user. Using our script will eliminate these limitations and ensures that the correct data are being imported no matter the structure of the Excel file provided. The importing Python script utilizes several libraries including "Pandas", "NumPy", and "re" (i.e., regular expression) to read the Excel files, parse its data into relevant relations, validate the type of the data that is being imported, and eventually create a backup of the database before adding the data into the database.

*3) Data Visualization*

A Python script was developed to connect to the database and perform data analysis and visualizations. The database was queried by the script with the help of "PyMySQL" library and the result were then further manipulated and plotted using "Pandas", "NumPy" and "Matplotlib".

## III. RESULTS

The designed database holds 8 relations and 164 attributes in total, for each patient. It currently stores data for 500 unidentifiable patients and will continue to grow as new patient's data become available. The designed Entity Relationship (ER) diagram is seen in Fig. 1, which resulted in the Relational Database shown in Fig. 2 and Fig. 3.

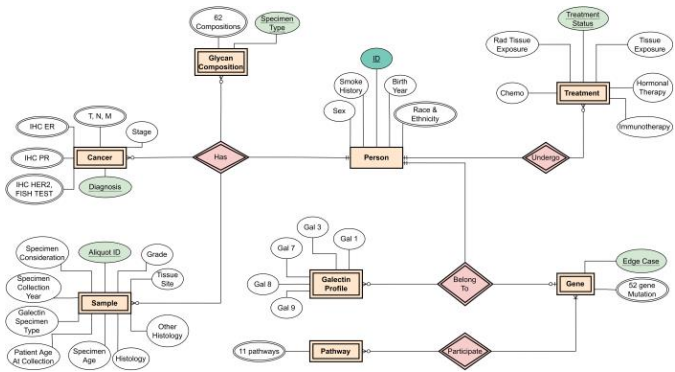

Fig. 1. Entity Relationship (ER) diagram

### A. Data Statistics

The dispersion of patient's birth year over sex is shown in Fig. 4, and it is noteworthy that the patient age distribution included as early as year 1920 and included year 2020 (2 years of age as of the writing of this paper). with the most data collected for females born in 1948. The patient's cancer type is presented in Fig. 5, and as most of the patients are females, additional data including males is being imported into the database. The patients' smoking history is presented in Fig. 6 which is dominated by females especially for the "Never smoked" category. In the following Fig. 7-9, cancer data distribution is shown against cancer stage and smoking history.

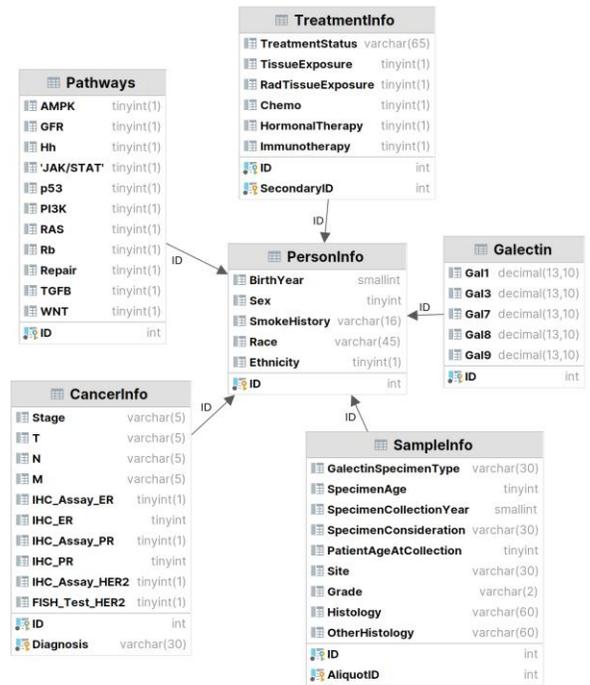

Fig. 2. The database schema showing primary and foreign keys

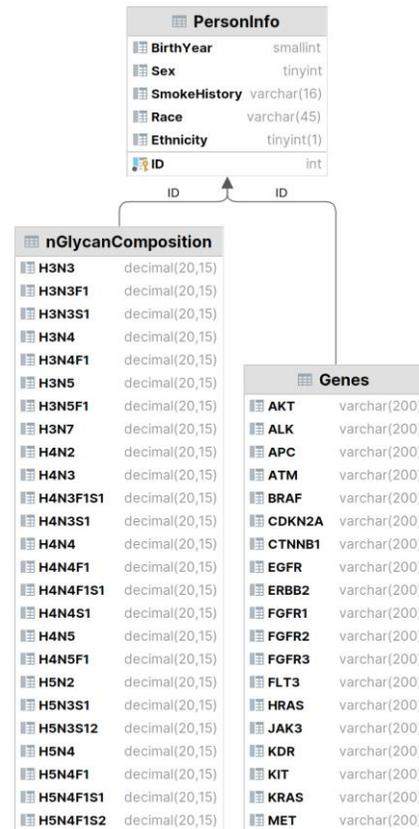

Fig. 3. The database schema showing primary and foreign keys for Genes and nGlycanComposition

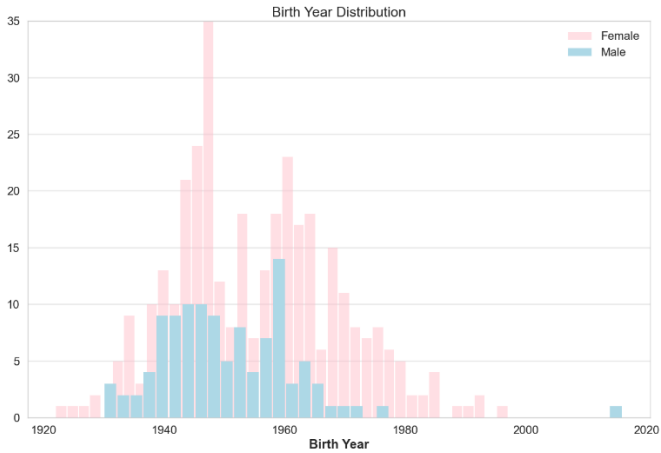

Fig. 4. Birth year data distribution categorized based on sex.

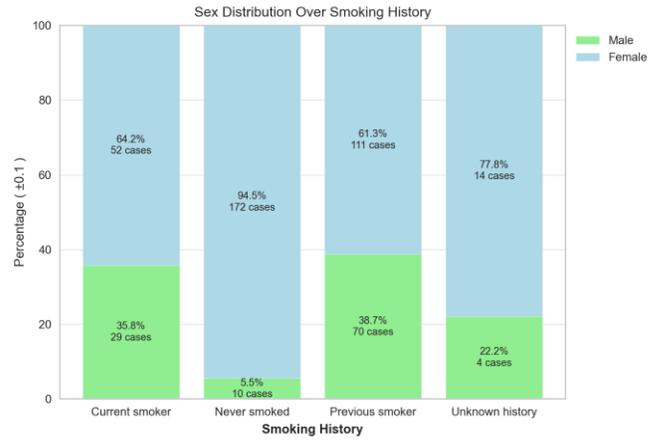

Fig. 6. Smoking history percentages categorized by sex

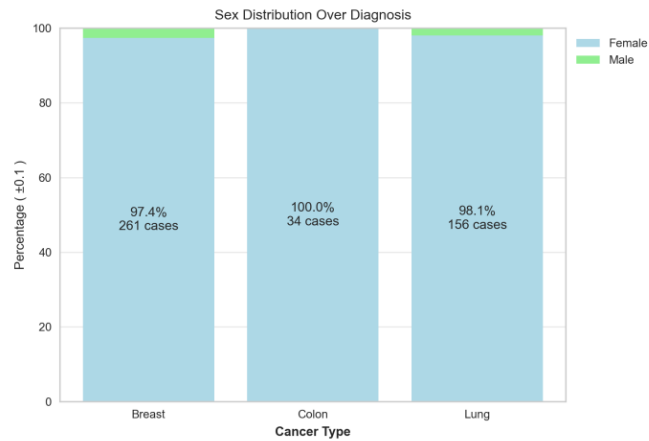

Fig. 5. Cancer types categorized based on sex.

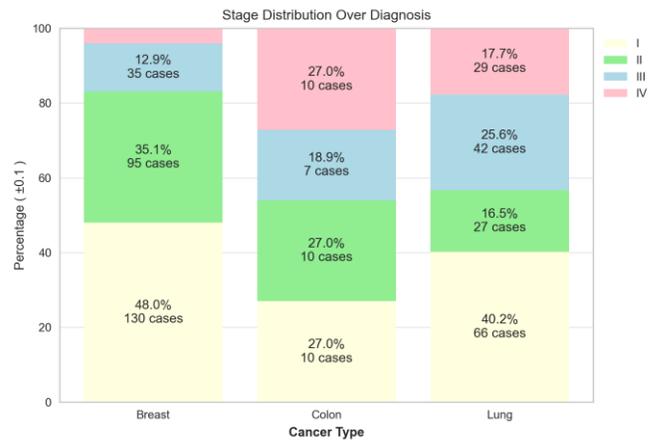

Fig. 7. Cancer type distribution by stage

## B. Data Visualization

Unison of MySQL and Python enabled the organization of all patients' information, their relative cancer information; the generation of glycomic, galectin, genetic, and lifestyle profiles.

Our interactive database allows providers to amalgamate cohorts from different groups to find correlations between different data types with the possibility of, for example, describing a "stage signature" based upon a combination of glycan signatures, galectin serum levels, genetic mutations, and lifestyle choices to create among our patient population.

Fig. 10-12 illustrate the database's use in searching for specific molecular analysis information. Fig. 10 demonstrates evaluation of patient's glycomic data by cancer stage. Fig. 11 shows mutation frequencies of a gene of interest while Fig. 12 displays serum levels of a protein that is under significant investigation in the oncological sphere.

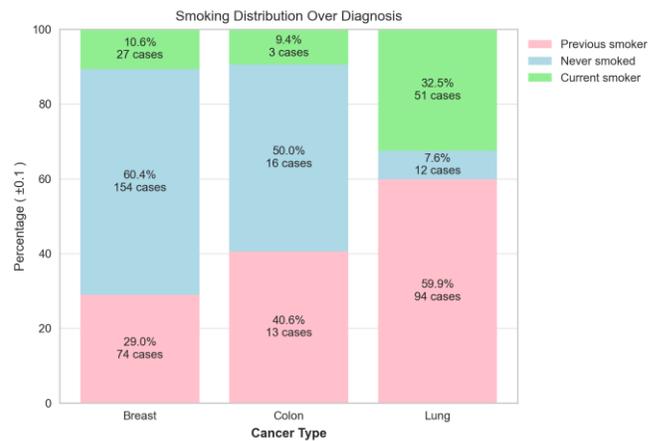

Fig. 8. Cancer type distribution by smoking history

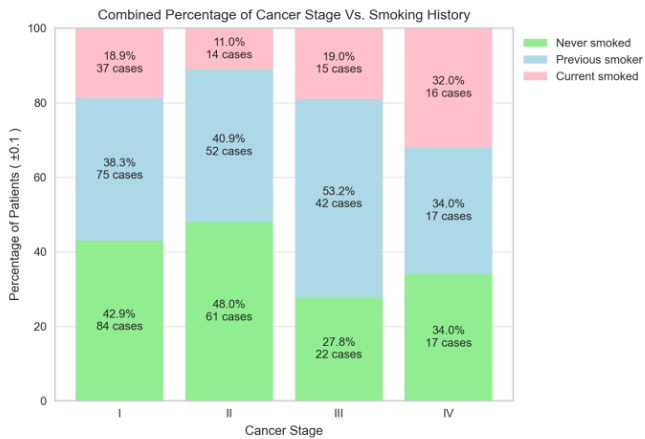

Fig. 9. Smoking history data distribution by cancer stage

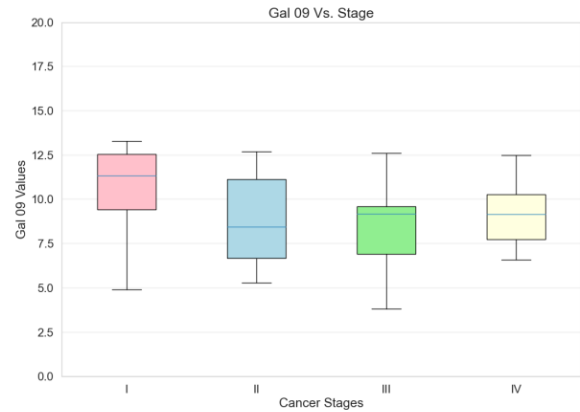

Fig. 12. Patient galectin profiling data. Example of results of database search by serum concentration of galectin-9 and breast cancer stage (I-IV)

## IV. DISCUSSION AND CONCLUSIONS

This work has clearly demonstrated the need for a centralized point of data access that combines different sources of data. Therefore, several databases were merged into a single access point to help ease the accessibility and use of the data. Furthermore, a 2000 line python script was developed to check, extract, and verify the integrity of data before importing to the database. Our interactive database allows providers to amalgamate cohorts from these groups to find correlations between different data types with the possibility of, for example, describing a "stage signature" based upon a combination of glycan signatures, galectin serum levels, genetic mutations, and lifestyle choices to create among our patient population.

In addition, the interface to the database was facilitated through an intuitive web application that serves as a database administration tool. We have also successfully integrated various types of molecular data into an interactive cancer patient database.

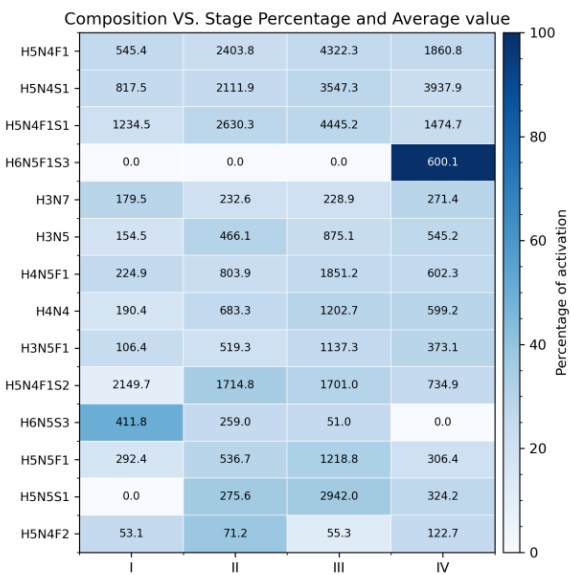

Fig. 10. Percent of patients with complex glycan structure present in serum in addition to their average value by cancer stage (I-IV)

Such databases will enable new discoveries through the incorporation of data analytics techniques. We believe the utility, such as the ability to analyze patterns (signatures) across stages in our database, will provide an opportunity for increased diagnostic and prognostic power. In addition, these molecular signatures may have practical application as a non-invasive diagnostic tool for tumor stage refinement. The integrated data in the form of a database will usher the use of artificial intelligence (AI) and Machine Learning in patient diagnostics and optimal treatment [39–42]. AI and ML techniques have been incorporated widely in the variety of domains from hardware to software and yielded promising results specially to improve healthcare. They have been used to automate hardware design [43], medical image segmentation [44], identification of the optimal treatment [45, 46], and maximizing the impact of drug dosage administration [47]. When combined with smart and wearable devices [40–42], a full picture of a person's health can be obtained to improve the quality of life and improvement of patient outcome.

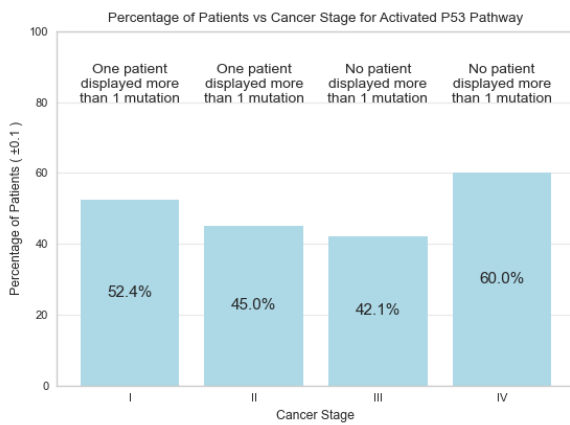

Fig. 11. Example of *TP53* gene mutation frequency

ACKNOWLEDGMENT

Research reported in this publication was supported by a grant from the National Institutes of Health (5R01LM011648 and P20 RR-01646100). The authors would also like to thank Basil Chaballout for contributing to the project.